\documentclass[twocolumn]{aa}
\usepackage{graphicx}
\usepackage{txfonts}
\input{amssym.def}
\input{amssym.tex}

\newcommand{\pcmcub}{\mbox{${\rm cm^{-3}}$}}

\newcommand{\pcmsq}{\mbox{${\rm cm^{-2}}$}}
\newcommand{\pwr}[2]{\mbox{$#1 \times 10^{#2}$}}

\newcommand{\UVsurf}{erg$\,$cm$^{-2}\,$s$^{-1}\,$\AA$^{-1}\,$arcsec$^{-2}$}

\newcommand{\lsim}{\mbox{$\mathrel{\vcenter{\hbox{\ooalign{\raise3pt\hbox{$<$}\crcr \lower3pt\hbox{$\sim$}}}}}$}}
\newcommand{\gsim}{\mbox{$\mathrel{\vcenter{\hbox{\ooalign{\raise3pt\hbox{$>$}\crcr \lower3pt\hbox{$\sim$}}}}}$}}

\def\mg2{Mg$_2$}

\def\kms{\relax \ifmmode {\,\rm km\,s}^{-1}\else \,km\,s$^{-1}$\fi}
\def\kkms{\relax \ifmmode {\,\rm{K\,km\,s}}^{-1}\else \,K\,km\,s$^{-1}$\fi}
\def\ha{\relax \ifmmode {\rm H}\alpha\else H$\alpha$\fi}
\def\hb{\relax \ifmmode {\rm H}\beta\else H$\beta$\fi}
\def\hi{\relax \ifmmode {\rm H\,{\sc i}}\else H\,{\sc i}\fi}
\def\hii{\relax \ifmmode {\rm H\,{\sc ii}}\else H\,{\sc ii}\fi}
\def\h2{\relax \ifmmode {\rm H}_2\else H$_2$\fi}
\def\lha{\relax \ifmmode L_{{\rm H}\alpha}\else $L_{{\rm H}\alpha}$\fi}
\def\shi{\relax \ifmmode \sigma_{{\rm HI}}\else $\sigma_{\rm HI}$\fi}   
\def\sh2{\relax \ifmmode \sigma_{{\rm H}_2}\else $\sigma_{{\rm H}_2}$\fi}
\def\degr{\hbox{$^\circ$}}
\def\arcmin{\hbox{$^\prime$}}
\def\arcsec{\hbox{$^{\prime\prime}$}}
\def\deg{\hbox{$^\circ$}}

\def\sec{\hbox{$^{\prime\prime}$}}
\def\fdg{\hbox{$.\!\!^\circ$}}
\def\fs{\hbox{$.\!\!^{\rm s}$}}
\def\farcm{\hbox{$.\mkern-4mu^\prime$}}
\def\farcs{\hbox{$.\!\!^{\prime\prime}$}}
\def\degd#1.#2{ #1\fdg#2 }                 
\def\mind#1.#2{ #1\farcm#2 }               
\def\secd#1.#2{ #1\farcs#2 }               
\def\hhh{\ifmmode {\rm ^h}              
         \else {${\rm ^h}$}
         \fi}
\def\sss{\ifmmode {\rm ^s}              
         \else {${\rm ^s}$}
         \fi}
\def\hms#1h#2m#3s{                      
                  \relax
                  \ifmmode #1^{\rm h}\,#2^{\rm m}\,#3^{\rm s}
                  \else \hbox{$#1^{\rm h}\,#2^{\rm m}\,#3^{\rm s}$}
                  \fi
                 }
\def\dms#1d#2m#3s{                      
                  \relax
                  #1\degr\,#2\arcmin\,#3\arcsec
                 }
\def\hmsd#1h#2m#3.#4s{                  
                      \relax
                      \ifmmode #1^{\rm h}\,#2^{\rm m}\,#3\fs#4
                      \else \hbox{$#1^{\rm h}\,#2^{\rm m}\,#3\fs#4$}
                      \fi
                     }
\def\dmsd#1d#2m#3.#4s{                  
                      \relax
                      #1\degr\,#2\arcmin\,#3\farcs#4
                     }
\def\mag{\relax                          
        \ifmmode ^{\rm m}
        \else $^{\rm m}$
        \fi
       }
\def\magd#1.#2{                          
              \relax
              \ifmmode #1^{\rm m}
                       \hskip-0.55em.\hskip0.22em#2
              \else \hbox{#1$^{\rm m}
                    \hskip-0.55em.\hskip0.22em$#2}
              \fi
             }

\input psfig.sty

\begin{document}

\title{{\it Research Note}\\ 
CO emission from candidate photo-dissociation regions in M81}

\author{J. H. Knapen\inst{1}
\and R. J. Allen\inst{2}
\and H. I. Heaton\inst{3}
\and N. Kuno\inst{4,5}
\and N. Nakai\inst{6}
}

\titlerunning{CO emission from candidate photo-dissociation regions in M81}
\authorrunning{J. H. Knapen et al.}

\offprints{J. H. Knapen}  

\institute{Centre for Astrophysics Research,
University of Hertfordshire, Hatfield, Herts AL10 9AB, UK\\
\email{j.knapen@star.herts.ac.uk} 
\and Space Telescope Science Institute, 3700 San Martin Drive,
Baltimore, MD 21218, USA
\and Johns Hopkins University, Applied Physics Laboratory, Laurel, 
MD 20723, USA
\and Nobeyama Radio Observatory, Nobeyama, Minamimaki, Minamisaku,
Nagano 384-13, Japan
\and The Graduate University for Advanced Studies (SOKENDAI),
2-21-1 Osawa, Mitaka, Tokyo 181-0015, Japan
\and Institute of Physics, University of Tsukuba, Tsukuba,
Ibaraki 305-8571, Japan}

\date{Received ; accepted 6 June 2006}

\abstract{

{\it Context} At least a fraction of the atomic hydrogen in spiral
galaxies is suspected to be the result of molecular hydrogen which has
been dissociated by radiation from massive stars.\\
{\it Aims} In this paper, we extend our earlier set of data from a
small region of the Western spiral arm of M81 with CO observations in
order to study the interplay between the radiation field and the
molecular and atomic hydrogen.\\
{\it Methods} We report CO(1-0) observations with the Nobeyama 45\,m
dish and the Owens Valley interferometer array of selected regions in
the Western spiral arm of M81.\\
{\it Results} From our Nobeyama data, we detect CO(1-0) emission at
several locations, coinciding spatially with \hi\ features near a
far-UV source. The levels and widths of the detected CO profiles are
consistent with the CO(1-0) emission that can be expected from several
large photo-dissociation regions with typical sizes of some $50 \times
150$\,pc located within our telescope beam. We do not detect emission
at other pointings, even though several of those are near far-UV sources
and accompanied by bright \hi. This non-detection is likely a
consequence of the marginal area filling factor of photo-dissociation
regions in our observations. We detect no emission in our Owens Valley
data, consistent with the low intensity of the CO emission detected in
that field by the Nobeyama dish. \\
{\it Conclusions} We explain the lack of CO(1-0) emission at positions
farther from far-UV sources as a consequence of insufficient heating and
excitation of the molecular gas at these positions, rather than as an
absence of molecular hydrogen.\\

\keywords{galaxies: individual: M81 -- galaxies: ISM --
ISM: molecules -- ISM: clouds -- radio lines: galaxies}

}

\maketitle

\section{Introduction}

Morphological evidence has been presented on several occasions over
the past twenty years that a non-negligible fraction of the atomic gas
found in the arms of spiral galaxies may be produced by the
destruction of molecular hydrogen in photo-dissociation regions
(PDRs) surrounding young massive stars.  This view, which was
initially developed by comparing the relative placement of dust lanes,
\hii\ regions, and \hi\ density enhancements in parts of M83 (Allen,
Atherton \& Tilanus 1985, 1986) and M51 (Tilanus \& Allen 1987), was
extensively tested by Allen et al. (1997, hereafter AKBS97) in M81,
based on the relative positions of a large number of far-UV
(FUV), \ha, and \hi\ sources. AKBS97 interpreted the morphological
information as implying that a significant part of the \hi\ in the
spiral arms of M81 is a product of star formation rather than a
precursor to it.

\begin{figure}
\centerline{\psfig{figure={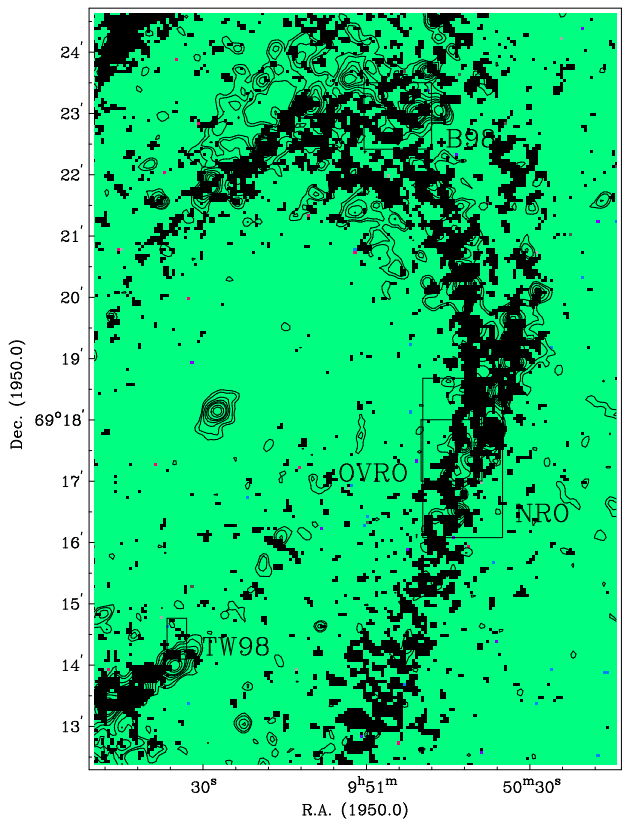},width=8.5cm,angle=0}}
\caption{FUV contours overlaid on a grey-scale representation of the
\hi\ integrated intensity map of the Western part of the inner disk of
M81. Overlaid are  the fields of our NRO and OVRO observations,
as well as the fields observed by Brouillet et al. (1998; labelled
B98), and by Taylor \& Wilson (1998; labelled TW98). Contour levels for
the FUV are $2.3, 3.1, 4.6, 6.2, 7.7, 9.2, 13.9, 18.5$ and
\pwr{36.9}{-18}\,\UVsurf.  Grey levels for the \hi\ vary linearly from
$1.5$ (light) to \pwr{7.3}{21}\,atoms \pcmsq\ (dark). FUV and \hi\ data
from AKBS97. }
\label{overview}
\end{figure}

Nevertheless, by conventional measures, the amount of molecular
hydrogen in M81 is not extensive.  For example, CO observations from
selected regions of M81, the locations of some of which are indicated
in Fig.~\ref{overview}, have been reported by Brouillet, Baudry \&
Combes (1988), Sage \& Westpfahl (1991) and Brouillet et al. (1991,
hereafter B91), who find relatively weak CO(1-0) emission from the
spiral arms on the basis of relatively low resolution
single-dish observations. Taylor \& Wilson (1998) reported the
detection of resolved emission from three molecular clouds in one of
those (Southern) regions with diameters of $\sim$100\,pc, and
Brouillet et al. (1998, hereafter B98) describe several clouds that
they found interferometrically in one of the (Northern) regions where
they had previously detected CO with a single dish.  These data
indicate that either the earlier hypothesis regarding the origin of
\hi\ gas is incorrect for M81, or there must be a reservoir of (cold)
molecular gas that has escaped detection.

We have obtained new single-dish and interferometric measurements in
the CO (1-0) line from a region that extends below the southernmost
field in the Western arm of M81 where B91 detected CO emission, to
determine whether evidence exists for a non-radiating molecular
reservoir.  The fields that we observed formed part of our previous
study (AKBS97) and contain several candidate PDRs.  Although CO gas
spanning the \hi\ velocity discontinuity caused by density wave-driven
streaming of the gas (e.g., Visser 1980) was not detected from our
interferometric observations, we obtained a number of single-dish
detections near several \hii\ regions. We ascribe these detections to
FUV excitation in the PDRs surrounding several young massive stars,
but more spatially resolved data will be required at those sites to
confirm this hypothesis.  While our results appear to corroborate the
absence of molecular gas at or just beyond the spiral arm shock front,
we argue that a population of large CO clouds similar to the cold
Galactic PDR observed at $l^{\rm II}\approx216\deg$ by Williams \&
Maddalena (1996, hereafter WM96) in the plane of the outer Galaxy
(i.e., G216$-$2.5) could exist at that location without being detected
at the best synthesised spatial resolution or single-dish sensitivity
achieved by our measurements.

\section{Empirical modelling of atomic and molecular volume densities}

Newly-formed massive stars produce copious amounts of FUV
emission. This flux of FUV photons interacts with the surface of the
parent molecular cloud and strongly affects its physics and chemistry
by heating, dissociating, and exciting the gas (see Hollenbach \&
Tielens 1999 for an excellent review of the properties of such
PDRs). When the presence of a PDR can be confirmed, e.g., from the
expected ``layered'' morphology, useful information can be derived on
the physical state of the gas from observations of the spectral lines
emitted by the excited atoms and molecules in those regions. Of the
many atomic constituents produced by photo-dissociation at and near
the surfaces of molecular clouds, atomic hydrogen (\hi) is among the
easiest to observe.
	
The PDRs found previously in M81 by AKBS97 manifest themselves as
bright ``arcs'' and ``blisters'' of enhanced 21\,cm \hi\ emission, located
near (and in some cases surrounding) sources of FUV continuum
emission.  Since the FUV emission is not always accompanied by bright
\ha\ emission, the sources of the dissociating photons must include
(and may even be dominated by) B stars. This picture has been
developed in more detail by Smith et al. (2000) who isolated 35
candidate PDRs in the giant spiral galaxy M101 and measured their \hi\
and FUV properties. A simple quantitative model for the \hi\
production in PDRs on the surfaces of molecular clouds was then used
to determine the total volume density of the gas in the underlying
molecular regions, which was found to lie in the range $30-1000\,{\rm
cm}^{-3}$, without a radial dependence of the mean value.

The production of \hi\ in PDRs by photodissociation has recently been
examined in more detail by Allen, Heaton \& Kaufman (2004, hereafter
AHK04) who confirmed the applicability of a simple analytic formula
(after some tuning of its parameters) initially proposed by Sternberg
(1988). In addition, AHK04 extended the picture to include CO(1-0)
emission, which is also produced in the same PDRs by the heating of
the ISM (mainly by grains via the photoelectric effect). AHK04 showed
that observations of the \hi\ column density and the CO(1-0) line
intensity of the same PDR can be combined into a diagnostic which
allows one to measure not only the density of the underlying molecular
cloud, but also the FUV flux incident upon it. They applied their
approach to G216$-$2.5, and concluded that an incident FUV flux of
$\sim$0.8 (in units of the local FUV flux in the ISM near the sun) was
bathing the surface of a large molecular cloud of density
$\sim$200\,cm$^{-3}$, thereby creating the observed levels of \hi\
column density and CO(1-0) surface brightness.
	
Since the size of the Galactic PDR observed by WM96
($100-200$\,pc) is of the same order as the linear resolution of
the earlier M81 observations analysed by AKBS97 ($\sim$150\,pc), we
choose that object as a template for estimating the level of CO(1-0)
emission we might expect in our targeted region of the spiral arms of
M81.  The total CO(1-0) emission from G216$-$2.5 is
7.1\,K\,km\,s$^{-1}$ (as quoted by AHK04), and from Fig.~1 of WM96, we
estimate that the total CO(1-0) profile width (FWHM) will be about
10\,km\,s$^{-1}$. From Fig.~2 in WM96, however, we estimate the extent
of this CO(1-0) emission to be $\sim$$50\times150$\,pc$^2$.  Since
our new single dish detections in M81 make use of a millimetre radio
telescope with an angular resolution of 16\,arcsec\,$\approx$\,280\,pc
(1\,arcsec\,=\,17.5\,pc at the 3.6\,Mpc distance we adopted for M81;
Freedman et al. 1994), beam dilution would reduce the expected total
emission to $\sim$0.86\,K\,km\,s$^{-1}$.

We have not attempted to compare this extrapolation with predictions
from the simple model developed in AHK04 of the integrated CO(1-0)
intensity expected near the shock front in M81 since that model
includes photon heating only.  The known correlation between the
velocity-integrated CO intensity and the nonthermal radio continuum
surface brightness in nearby galaxy disks (Adler, Allen, \& Lo 1991)
and in the Galaxy (Allen 1992) suggests that an enhanced level of
cosmic rays also has something to do with heating the ISM, and
therefore with elevating the CO surface brightness.  The AHK04 model
also utilizes the microturbulent velocity width of a single cloud
rather than the macroturbulent width typifying an extragalactic
ensemble.  Nevertheless, the faintness of nonthermal radio continuum
emission from M81 indicates that it is a better choice for assessing
the issues under investigation than other nonthermally-bright nearby
galaxies such as M51, M83, and NGC~6946 because the reduced role of
cosmic ray heating provides a more controlled experiment.

\section{Observations, data handling, and results}

\subsection{Nobeyama data}

Observations of the $^{12}$CO($J$ = 1 -- 0) emission were made during
1998 January 23 to 26 and March 27 to 31 with the 45\,m telescope at
the Nobeyama Radio Observatory (NRO).  At the rest frequency of the
$^{12}$CO(1-0) transition (115.271204\,GHz), the half-power beam width
(HPBW) is 16\,arcsec. The aperture and main-beam efficiencies were
$\eta_{\rm a} =0.29$ and $\eta_{\rm mb} =0.40$, respectively.

We used the $2\times2$ Semiconductor - Insulator - Semiconductor (SIS)
focal-plane array receiver which can simultaneously observe four
positions separated on the sky by 34\,arcsec each (Sunada et
al. 1995). Differences in $T_{\rm A}^*$ between the four beams due to
the different sideband ratios were adjusted to be equal. We used the
SIS single beam receiver with a Martin-Puplett type SSB filter for an
image sideband termination at the cryogenic temperature (4\,K) to
observe selected points effectively. As receiver backends we used
2048 channel wide-band acousto-optical spectrometers (AOS). The
frequency resolution and channel spacing were 250\,kHz and 125\,kHz,
respectively. At 115\,GHz the corresponding velocity resolution and
velocity coverage are 0.65\,km\,s$^{-1}$ and
650\,km\,s$^{-1}$. Calibration of the line intensity was made by the
chopper-wheel method (Ulich \& Haas 1976), yielding the antenna
temperature $T_{\rm A}^*$ corrected for atmospheric attenuation. In
this paper, we use the main beam brightness temperature $T_{\rm
mb}\equiv T_{\rm A}^*/\eta_{\rm mb}$ as the intensity scale of the CO
brightness temperature. The system noise temperature (SSB), including
the atmospheric effect and the antenna ohmic loss, was 500 -- 800\,K in
$T_{\rm A}^*$, depending on the elevation of the source and the
atmospheric conditions. We smoothed the spectra to a velocity
resolution of 10\,\kms.

We observed 46 points in a region of about $70\times90$\,arcsec. The
separation of each point is 11\,arcsec in right ascension and
declination. The telescope pointing was checked and calibrated every
few hours by observing the SiO maser emission of the late-type star
R-UMa at 43\,GHz. The absolute pointing accuracy was better than
5\,arcsec (peak value) throughout the observations.

At the data reduction stage, we checked all individual spectra, with
integration times of 20\,s, and eliminated those with bad
baselines. The remaining spectra were coadded and fitted with a linear
baseline.

\begin{figure*}
\centerline{\psfig{figure={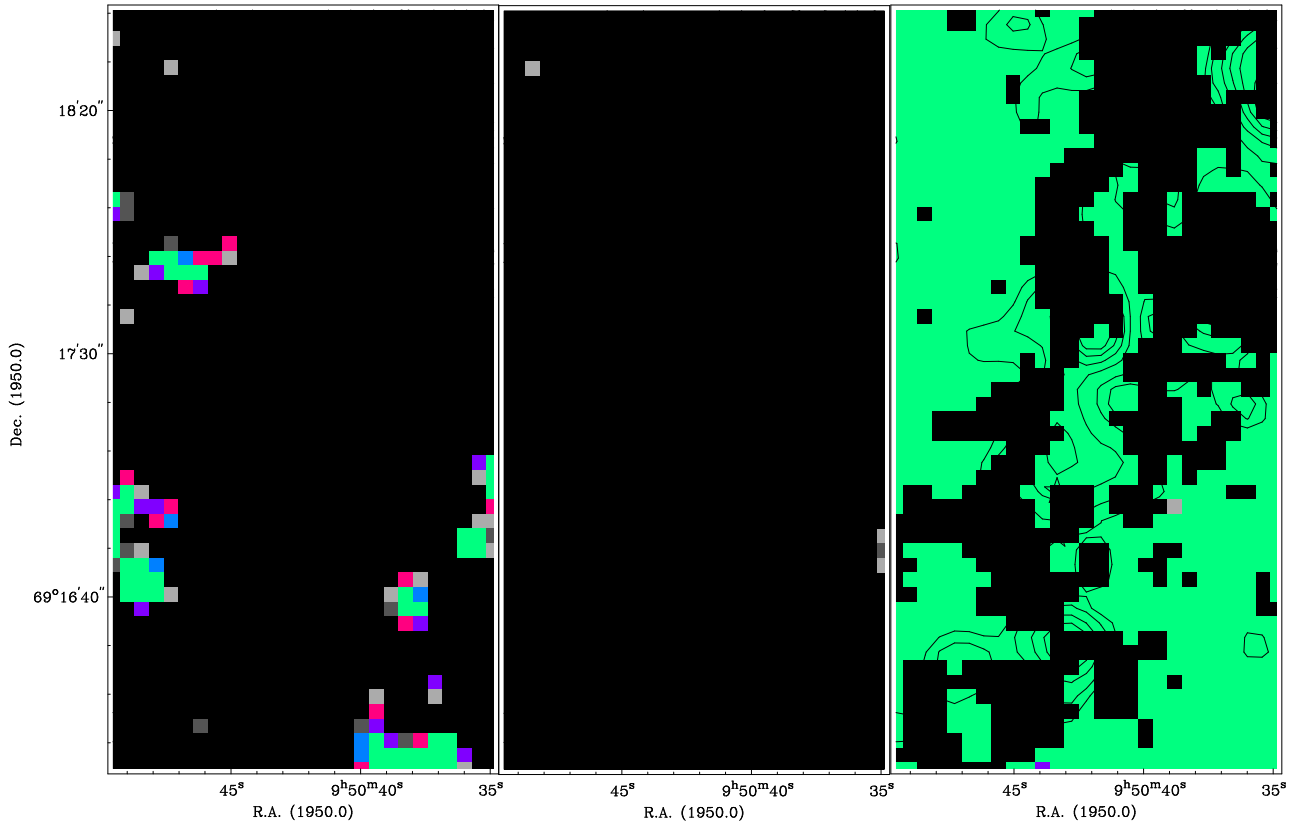},width=\textwidth,angle=270}}
\caption{The region of the spiral arm in M81 studied in the present
paper. Contours in all three panels are of the FUV emission, and
contour levels are $2.3, 3.1, 4.6, 6.2, 7.7, 9.2, 13.9, 18.5$ and
$36.9\times10^{-18}$
erg$\,$cm$^{-2}\,$s$^{-1}\,$\AA$^{-1}\,$arcsec$^{-2}$.  The contours
are overlaid on grey scale representations of the FUV (left panel,
grey levels from 0.8 [light] to $15\times10^{-18}$
erg$\,$cm$^{-2}\,$s$^{-1}\,$\AA$^{-1}\,$arcsec$^{-2}$ [dark]);
H$\alpha$ (middle panel; grey levels from 0.5 to $95\times10^{-17}$
erg$\,$cm$^{-2}\,$s$^{-1}\,$arcsec$^{-2}$); and \hi\ (right panel,
grey levels from $1.5$ to $7.3\times10^{21}$\,atoms cm$^{-2}$). The
positions of our three most credible CO detections from our NRO
observations (B, C, and D in Fig.~\ref{pointings}) are indicated in
each panel by open star symbols.  Centre positions of the dust
filaments as measured by Kaufman et al. (1989a) are indicated by the
line segments drawn in the middle panel.}
\label{region}
\end{figure*}

\begin{figure}
\centerline{\psfig{figure={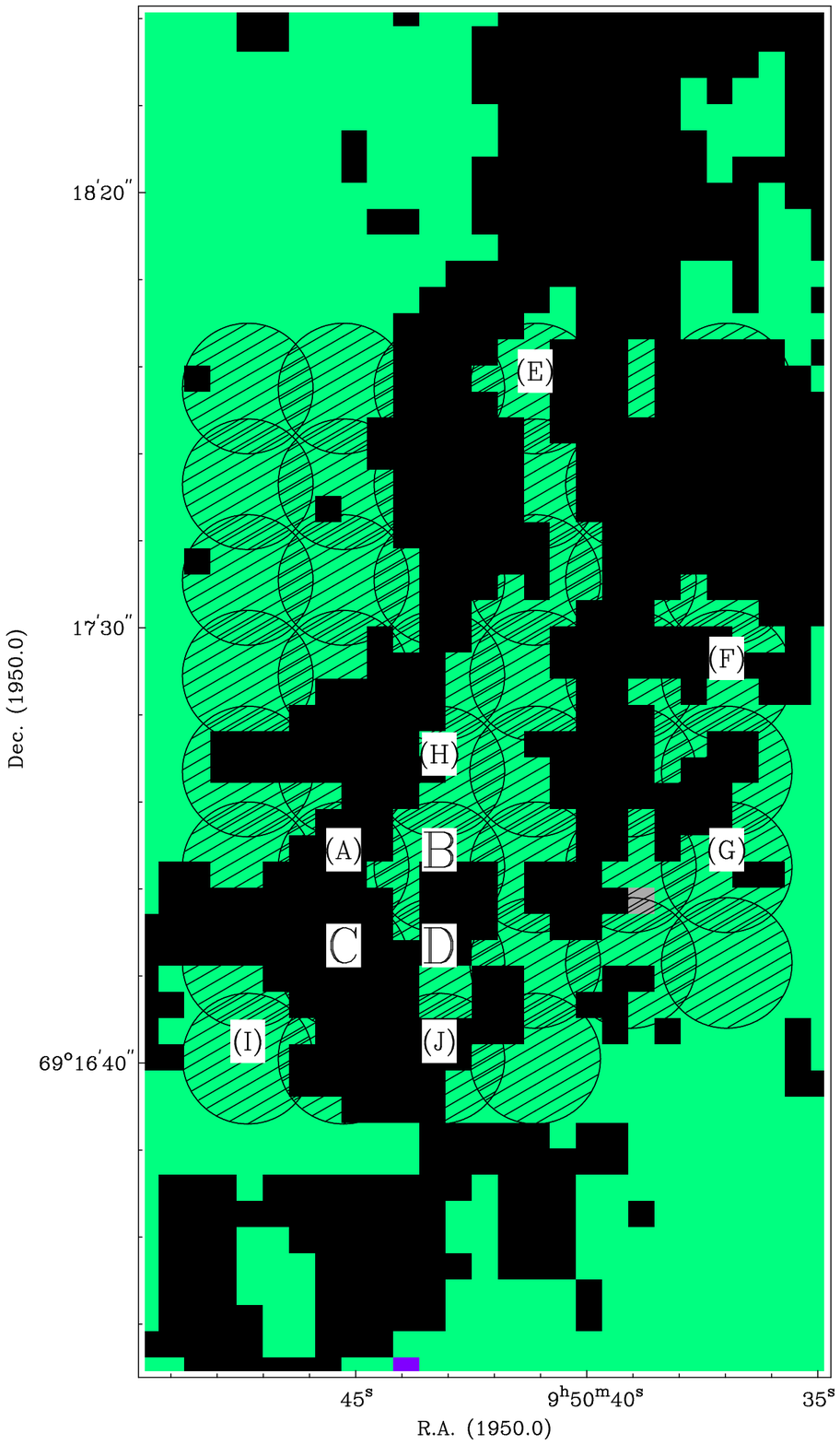},width=0.45\textwidth,angle=0}}
\caption{Positions of the NRO 45\,m beam pointings, overlaid on the
\hi\ map as shown in the right panel of Fig.~\ref{region}. Grey levels
as in Fig.~\ref{region}. Letters denote positions where CO was
detected (see text); less secure detections are denoted by letters in
parentheses.}
\label{pointings}
\end{figure}

The region of interest is shown in detail in Fig.~\ref{region}, where
we show the FUV, \ha, and \hi\ distributions at 9\,arcsec resolution in
the same format as used in AKBS97 for five larger regions of M81. Dust
lanes (based on positions given by Kaufman, Elmegreen \& Bash 1989a)
are also indicated, in the middle panel. Figure~\ref{pointings} shows
the NRO 45\,m beam positions overlaid on the grey scale representation
of the \hi\ data, as shown in the right panel of Fig.~\ref{region}.
The beam positions designated with letters correspond to detections,
as discussed in more detail below, and as tabulated in
Table~\ref{data}.

\begin{figure}
\vspace{1cm}
\centerline{\psfig{figure={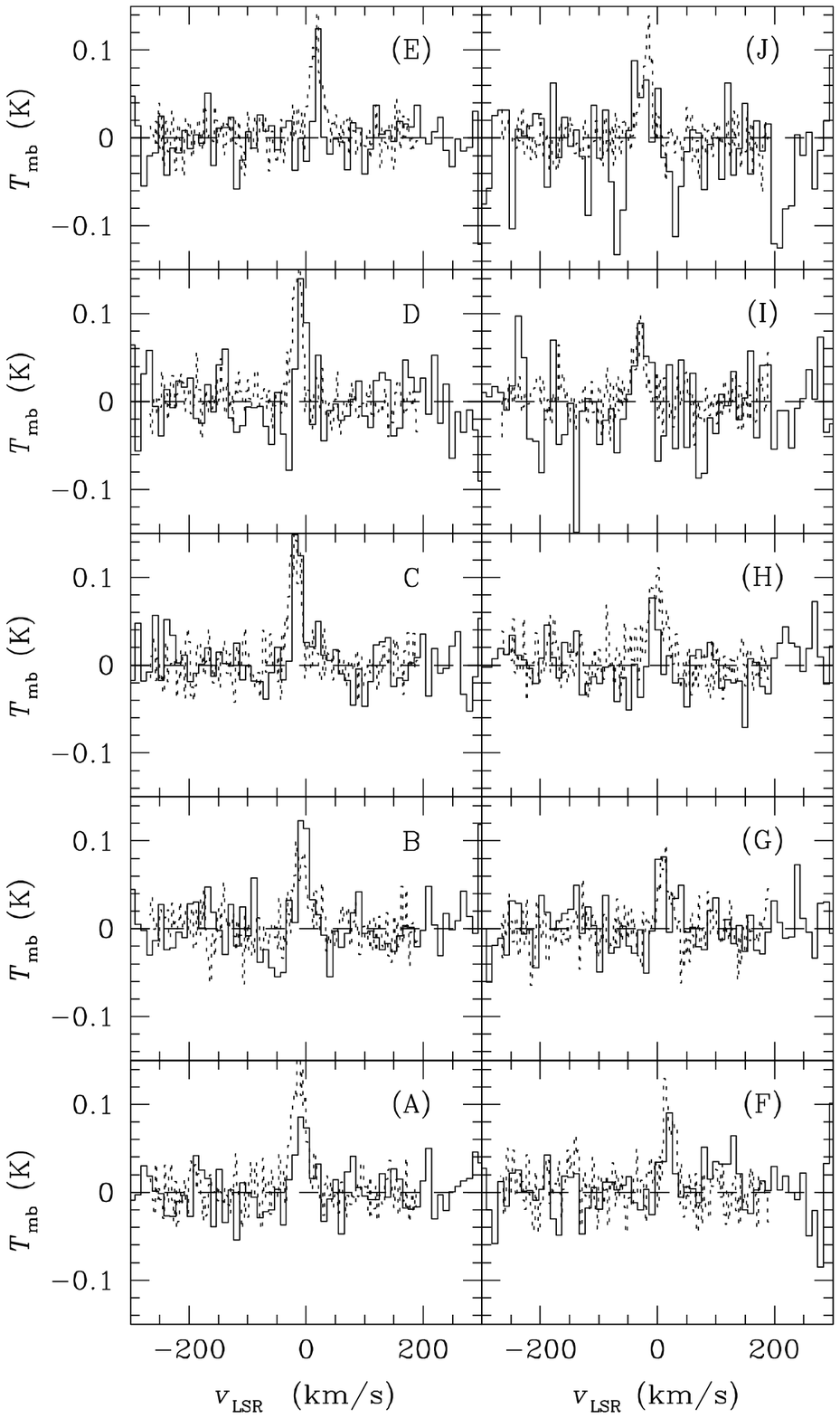},width=9cm,angle=0}}
\caption{Final CO(1-0) spectra from NRO (drawn lines), smoothed
to $dv=10$\,\kms\ velocity resolution. Letters in this figure
correspond to positions of the telescope pointing as shown in Table~1
and Fig.~\ref{pointings}, with the (0,0) offset corresponding to
position RA=\hms 09h 50m 44.0s, Dec=\dms 69d 17m 30s (B1950). They
indicate spectra discussed in more detail in the paper; less
secure detections are denoted by letters in parentheses. The dotted
lines indicate \hi\ spectra, on an arbitrary scale, at the same
position.}
\label{spectra}
\end{figure}

\begin{table}[htb]
\begin{center}
\begin{tabular}{lccccc}
\hline\noalign{\smallskip}
Pos. & Offset & $\int T_{\rm mb}\delta v$ & $v_{\rm peak}$ & d$v$
& $T_{\rm mb, peak}$ \\
        & (\sec) & (\kkms) & (\kms) & (\kms) & (mK) \\
\noalign{\smallskip}
\hline
\noalign{\smallskip}
A & 5.5 $-$27.5	       & 2.32 & $-6.3$  & 25.3 & 90  $\pm$ 24 (3.7)\\
B & $-$5.5 $-$27.5     & 2.73 & $-4.0$  & 18.2 & 145 $\pm$ 23 (6.3)\\
C & 5.5 $-$38.5	       & 2.52 & $-14.6$ & 14.7 & 187 $\pm$ 24 (7.8)\\
D & $-$5.5 $-$38.5     & 1.86 & $-6.6$  & 15.6 & 155 $\pm$ 32 (4.8)\\
E & $-$16.5 27.5       & 1.20 & 20.6    & 7.7  & 125 $\pm$ 24 (5.2)\\
F & $-$38.5 $-$5.5     & 1.40 & 20.6    & 14.3 & 87  $\pm$ 25 (3.5)\\
G & $-$38.5 $-$27.5    & 1.47 & 10.5    & 14.4 & 85  $\pm$ 26 (3.2)\\
H & $-$5.5 $-$16.5     & 1.07 & $-9.1$  &  8.6 & 75  $\pm$ 23 (3.3)\\
I & 16.5 $-$49.5       & 2.50 & $-29.0$ & 19.5 & 100 $\pm$ 45 (2.2)\\
J & $-$5.5 $-$49.5     & 3.19 & $-38.8$ & 26.4 & 90  $\pm$ 46 (2.0)\\
\hline
\end{tabular}
\end{center}
\caption[ ]{CO(1-0) measurements obtained using the 45\,m NRO
dish, $dv=10$\,km\,s$^{-1}$, $T_{\rm mb}=T^*_{\rm A}/0.40$. The
identification of the measurement, alphabetically as used elsewhere in
this paper, and as a position offset from (0,0), which is RA=\hms 09h
50m 44.0s, Dec=\dms 69d 17m 30s (B1950), is given in cols.~1 and 2;
the integrated intensity in K\,\kms\ is in col.~3; the peak position
and width of the profile, corresponding to velocity and velocity
dispersion, are given in cols.~4 and 5; and the peak main beam
temperature is in col.~6, where the error quoted is the rms of the
spectrum (the number in parentheses is the ratio of the two, hence
spectrum A is considered a $90/24=3.7\sigma$ detection).  }
\label{data}
\end{table}

A sample of the final CO(1-0) spectra is shown in
Fig.~\ref{spectra}. At most of the 46 positions we did not detect any
emission, which corresponds to upper limits ranging from 48 to 201\,mK
(3\,$\sigma$). We did, however, detect CO emission at some of the
positions, which are indicated by letters in Figs.~\ref{pointings}
and~\ref{spectra}. The \hi\ spectra shown in Fig.~\ref{spectra},
derived from a data set kindly made available by E.~Brinks \&
F.~Walter (2005, private communication), show that in all cases our detected
CO emission is at the expected velocity. The most convincing of our
detections are labelled B, C, and D ($\gtrsim5\sigma$ detections in terms of
$T_{\rm mb, peak}$; see Table~\ref{data}). While E also satisfies this
5$\sigma$ detection criterion, it spans only a single channel on
Fig.~4, and is therefore less convincing. The detections B, C, and D
are clustered around a weak peak in the FUV emission at
RA=\hms09h50m43s, Dec=\dms69d17m10s (B1950.0), and are located close
to the \hi\ arc which almost completely surrounds this FUV
peak. Around this cluster of detections there are four other, more
tentative, detections, where an emission peak is seen at the expected
velocity, but with a peak main beam brightness of only $2-4\sigma$ (A,
H, I, J).  At two further positions, labelled F and G, we detect
emission with a peak main beam brightness of $\sim$$3\sigma$. Peak
antenna temperatures, r.m.s. noise levels, and results from Gaussian
fits for all the pointings discussed above are summarized in
Table~\ref{data}.

\subsection{Owens Valley data}

The millimetre array at the Owens Valley Radio Observatory (OVRO) was
used on two separate occasions in its low spatial resolution mode to
make CO(1-0) observations toward a region in the Western arm of M81
(see Fig.~\ref{overview}) which corresponds closely to the $1\times
1$\,arcmin box shown by B91 to the SE of their region N1.  The minimum
and maximum baselines were 15\,m and 115\,m during both sessions,
corresponding approximately to spatial scales of 620 and 80\,pc,
respectively, at the adopted distance to M81.  Tracks were obtained on
1998 May 22-23 and 1998 November 5 comprising repetitive cycles of five
separate 5\,minute integrations on-target and two 5\,minute integrations
on the gain calibrator. Since the phase centre for the target
observations was RA=\hms 09h 50m 44.0s, Dec=\dms 69d 17m 30s (B1950),
our 62.9\,arcsec primary beam fell nearly completely within the region
observed with the NRO dish (Fig.~\ref{overview}).

The digital correlators at Owens Valley were configured to provide two
slightly overlapping spectral bands in the upper sideband, each with
32 channels of 1 MHz (2.6\,\kms) width.  These bands were
subsequently truncated and merged to provide 56 contiguous channels
for each track.  The data for each track were then merged to produce a
dataset with 2480 visibilities corresponding to 13.78 hrs of on-source
integration with a visibility-weighted average system temperature of
989\,K.  The resulting velocity coverage, centred at $-30$\,\kms\
(relative to the local standard of rest), extended from $+42.8$ to
$-102.8$\,\kms.

All CO mapping was performed with the Astronomical Image Processing
System (AIPS) package, using uniform weighting to preserve any
extended faint emission in the field.  The data were averaged in
velocity to obtain a datacube with 28 separate 5.2\,\kms\
channels. The restoring beam was a Gaussian of width $3.7\times 3.6$
arcsec (FWHM) at a position angle of 8.2\,deg.  The flux density noise
in the dirty map was 0.032\,Jy\,beam$^{-1}$ (uncorrected for primary beam
attenuation) which equates to a temperature noise of 221\,mK in the
main beam.  In addition, the visibilities were tapered to also form a
$9\times 9$ arcsec beam at a position angle of 46.6\,deg for
comparison with the \hi, H$\alpha$, and FUV datasets analysed
previously by AKBS97.  The noise in the latter (uniformly weighted,
uncorrected) dirty map was 0.055\,Jy$/$beam$\equiv 62.4$\,mK.  Our
resulting maps are 16.4 and 6.7 times more spatially resolved,
respectively, than the single dish survey maps made in this region by
B91.  Since the latter had a sensitivity of $T_{\rm R}^\ast$ $\approx
20$\,mK and an on-target observation time of around 0.5\,hr, equation
C2 of AHK04 shows that our new 9 arcsec map has a flux density
sensitivity that is approximately 14 times better than the previous
survey data.

No clear evidence was obtained for CO emission to the limit of our
sensitivity, neither in our 5.2\kms\ channel maps prepared at
3.7\,arcsec, nor in those at 9\,arcsec resolution.  While several
features appear on each map, the number of features with positive flux
density exceeding preset thresholds roughly equals the number of
negative-valued threshold crossings.  When the smoothed (9\,arcsec)
flux density maps were filtered at 3$\sigma$, no features repeated
from one channel map to the next.

To further quantify these results, we have derived an upper limit to
the CO(1-0) flux density from this region based on noise estimates
made with the AIPS task {\sc imstat}, which were derived by fitting
Gaussian functions to pixel histograms. The corresponding 3$\sigma$
upper limit to the flux density value is 163.8\,mJy\,beam$^{-1}$.
Accordingly, the velocity-integrated 3$\sigma$ upper limit in our
field is (163.8\,mJy\,beam$^{-1})\times(2\times 10^6$\,Hz) $ =
3.276\times 10^5$\,(Jy\,beam$^{-1}$)\,Hz over the 2-channel span
corresponding to our chosen velocity resolution.  This corresponds to
0.85\,Jy\kms\ or 1.03\kkms\ for our 9\,arcsec beam. Although several
of our strongest detections at NRO had integrated intensities greater
than this level, only NRO pointings E, H, and part of B fell within
our primary beam at Owens Valley.  The integrated intensities from E
and H are near our OVRO sensitivity limit, and B is on the periphery
of the primary OVRO beam.

\section{Discussion}

The results described in the previous section from the large dish at
Nobeyama have extended the regions where CO has been found in the
Western arm of M81 to locations that are farther down that feature
than the southernmost field detected by B91.  While interesting in
itself, the central issue in interpreting this result is: why did the
emission stop there?  The two most plausible explanations are that the
lack of CO(1-0) emission indicates the absence of measurable
quantities of molecular gas in dark regions of the observed field, or
that the field contains low density gas that isn't heated sufficiently
to reveal its presence radiatively.  The first interpretation is a
consequence of the prevailing view that the CO(1-0) surface brightness
and the column density of molecular hydrogen are linearly related,
independent of any effects caused by localised heating (Young \&
Scoville 1991; Combes 1991).  In this case, the \hi\ ridge observed
downstream from the velocity shock front probably represents the
compression of previously existing atomic gas (Kaufman et al. 1989b).

An alternative explanation that is consistent with our observations is
that the low flux of cosmic rays and the low FUV surface brightness
are inadequate to sufficiently heat the ISM to shine in the CO(1-0)
line in most of our field. A low cosmic ray flux is implied by the
faint levels of nonthermal radio continuum emission from M81, and low
FUV levels are evident in many regions of Figure~2.  Emission is
observed in areas where heating by photons through the photoelectric
effect, and by cosmic rays through ionisation, leads to a rise in the
kinetic temperature to a level sufficient, along with the density, to
allow the radiating CO energy levels to be populated.

To illustrate these points, we note that the cosmic ray intensity
appropriate for the Galaxy in the neighbourhood of the Sun would heat
giant molecular clouds (GMCs) that are just dense enough to have
thermalized CO emission ($n\approx10^4$\,cm$^{-3}$) only to $\sim$5\,K
(Tielens 2005, p.~352). This is much lower than is observed in
localized regions of GMCs in the Galaxy. At densities below $\sim$$10^3$
the CO level populations would be subthermally excited, leading to
even lower radiation temperatures. As justification for comparing M81
with the Milky Way, we note that the radio continuum surface
brightness, which can be used as a proxy for the cosmic ray flux, is
similar in the disks of our Galaxy and M81, at a level that is an
order of magnitude lower than that in M51 (Adler et al. 1991; Allen
1992).

We can readily draw two conclusions from this discussion. First,
regions showing high-temperature CO emission must be heated by a
supplementary source in addition to cosmic rays -- this source is most
likely photoelectric heating of the gas caused by dust absorption of FUV
photons, although in some cases \h2\ formation heating may play a role
(R\"ollig et al. 2006). In Galactic GMCs, such sources can usually be easily
identified as nearby recently-formed massive O/B stars. Second, in the
absence of such photons and a sufficient level of cosmic ray flux, the
cooling in GMCs is so efficient that the kinetic temperature will
quickly fall below 5\,K, making the detection in any {\it emission}
line tracer virtually impossible. This, we postulate, is what happens
in M81.

The lack of CO emission at locations between the \hi\ velocity shock
and the atomic hydrogen ridge, and its weak detection near several
\hii\ regions downstream from the shock, is consistent with the
interpretation by Allen et al. (1985, 1986) and Tilanus \& Allen
(1987) of morphological data from M83 and M51, i.e., that the
interstellar gas remains primarily molecular after passing through the
shock until it dissociates in the neighbourhood of giant \hii\ region
complexes downstream.  The spatial extent of our observations is more
than adequate to search for such patterns. Since one arcsec
corresponds to $\sim$17.5\,pc at the 3.6\,Mpc distance assumed for
M81, our 63\,arcsec OVRO field alone spans nearly 1.1\,kpc.  This is
more than twice the separation claimed by Kaufman et al. (1989b) for
the distance between the \hi\ velocity shock and the atomic hydrogen
ridge in the Western arm of the latter galaxy.

The cold Galactic cloud described by WM96, G216$-$2.5, provides a
useful template for what we might observe from low density gas clouds
in M81.  Although AHK04 found that the most likely density for this
cloud was 200\,\pcmcub, the uncertainty in that determination was
large enough to render the role of collisional excitation uncertain.
Since the beam-diluted integrated intensity of this PDR at the
distance of M81 would be $\sim$0.9\kkms\ when observed by the
16\,arcsec NRO beam, similarly excited clouds with similar profile
widths ($\sim$10\kms, cf. Sect.~2) should be apparent in our
observations.  The three firm detections found earlier (i.e., B, C,
and D) are all somewhat brighter (by a factor of $2-3$) and
broader (by a factor of $\sim$1.5) than this estimate, perhaps
indicating the presence of multiple PDR structures at these positions.

The absence of CO emission in our single beam fields that cover the
region between the velocity front and the \hi\ ridge sets an upper
limit of $\sim$0.8\kkms\ for the integrated intensity of the gas from
that region.  If low density molecular gas does reside there, it is
not excited sufficiently by the low FUV flux to be detected at the
best sensitivity we achieved.  The lack of emission at more strongly
irradiated locations along the \hi\ ridges in Fig.~3 could easily be
caused by enhanced beam dilution from structures that are smaller in
linear size than the WM96 cloud.  The WM96 cloud would have a
beam-diluted integrated intensity $\sim$2.7\kkms\ in our 9\,arcsec
synthesized OVRO beam at the distance of M81.  Although this exceeds
the 3$\sigma$ sensitivity value determined for that array by a factor
of $\sim$2.6, the pointings detected by the Nobeyama telescope only
partially fell within the OVRO field.

To test the consistency of our findings with earlier results, and the
plausibility of using the WM96 PDR as a template for modelling putative
undetected gas clouds in our region of the Western arm, we
reinterpret the results obtained by B98 with the IRAM Plateau de
Bure interferometer.  The location of their field is shown in our
Fig. 1, while the resulting CO(1-0) map appears in Fig. 1 of their
paper.  From the extent and maximum intensity of the structures shown
in their figure, we estimate that the NRO 45\,m beam would have
reported a brightness of only $\sim$0.4\kkms\ for their region C3.
The combination of regions C1 and C2, which could fall into one of our
beams, would have an integrated intensity of about 1\kkms.  Thus, we
would not have detected clouds in our area with the characteristics
found by B98 for C3, while the combination of clouds C1 and C2 would
have  yielded only a marginal detection.  None of these clouds would
have been detected at the 3$\sigma$ level in our 9\,arcsec OVRO maps.

When we compare the extrapolated properties of the G216$-$2.5 PDR with
the earlier CO results from B98, it is clear that they have similar CO
brightnesses ($\sim$0.8\kkms\ for the latter) and sizes
(50$\times$100\,pc for the WM96 cloud, and at most $150-250$\,pc for the
B98 clouds).  The general lack of CO detections in the area observed
with the NRO dish is entirely consistent with this picture: we detect
a few stronger peaks, but fail to detect CO at the majority of our
pointings. This is exactly what we would expect to see if we had
observed clouds in our region that were similar to those found by B98
in their field: a marginal detection from C1$+$C2, and no detections
elsewhere in the field.  We thus ascribe our lack of CO detections to
being predominantly a filling factor effect --- individual CO-emitting
regions may be relatively bright, but in our 16\,arcsec NRO beam they
would be diluted to below the detection threshold.

\section{Conclusions}

We have detected CO(1-0) emission at several locations showing bright
\hi\ features near FUV sources in M81. The levels and widths of the
detected CO profiles are consistent with the CO(1-0) emission to be
expected from several large photo-dissociation regions with typical
sizes of $\sim$$50 \times 150$\,pc located within our NRO beam of size
280\,pc. Non-detections, either at other pointings near FUV sources
that contain bright \hi, or in our synthesized maps, are likely to be
a consequence of the marginal area filling factors of PDRs in our
observations, and future surveys should be prepared to substantially
increase the sensitivity and spatial resolution over what we have
achieved here. The lack of CO(1-0) emission at positions farther from
the sources of FUV is consistent with insufficient heating of the ISM,
resulting in insufficient excitation of the CO molecules.

\begin{acknowledgements}

We thank Dr. N. Brouillet for providing a CO sensitivity estimate
germane to the undetected field just to the South-East of their field
N1, and Prof. E.~Brinks and Dr. F.~Walter for providing access to
unpublished \hi\ data of M81.  RJA is grateful to Prof. J.~H. Hough at
the University of Hertfordshire and Dr. S. Beckwith at the Space
Telescope Science Institute for their generous travel support in
connection with the completion of this paper. JHK wishes to
acknowledge the Leverhulme Trust for the award of a Leverhulme
Research Fellowship, as well as the hospitality of our Japanese
colleagues during his visit to Nobeyama.

\end{acknowledgements}


\begin{thebibliography}{}

\bibitem[Adler et al.(1991)]{1991ApJ...382..475A} Adler, D.~S., Allen,
R.~J., \& Lo, K.~Y.\ 1991, \apj, 382, 475

\bibitem[Allen(1992)]{1992ApJ...399..573A} Allen, R.~J.\ 1992, \apj, 399, 
573 

\bibitem[Allen et al.(1985)]{1985bems.symp..243A} Allen, R.~J.,
Atherton, P.~D., \& Tilanus, R.~P.~J.\ 1985, in Birth and Evolution of
Massive Stars and Stellar Groups, eds. W. Boland \& H. van Woerden
(Dordrecht: Reidel), 243

\bibitem[Allen et al.(1986)]{1986Natur.319..296A} Allen, R.~J., Atherton, 
P.~D., \& Tilanus, R.~P.~J.\ 1986, \nat, 319, 296 
 
\bibitem[Allen et al.(2004)]{2004ApJ...608..314A} Allen, R.~J., Heaton, 
H.~I., \& Kaufman, M.~J.\ 2004, \apj, 608, 314 (AHK04)

\bibitem[Allen et al.(1997)]{1997ApJ...487..171A} Allen, R.~J., Knapen, 
J.~H., Bohlin, R., \& Stecher, T.~P.\ 1997, \apj, 487, 171 (AKBS97)

\bibitem[Brouillet et al.(1988)]{1988A&A...196L..17B} Brouillet, N., 
Baudry, A., \& Combes, F.\ 1988, \aap, 196, L17 

\bibitem[Brouillet et al., \ 1991]{bro91} Brouillet, N., Baudry, A.,
Combes, F., Kaufman, M., \& Bash, F. 1991,
 \aap, 242, 35 (B91)

\bibitem[Brouillet et al.(1998)]{1998A&A...333...92B} Brouillet, N.,
Kaufman, M., Combes, F., Baudry, A., \& Bash, F.\ 1998, \aap, 333, 92
(B98)

\bibitem[Combes(1991)]{1991ARA&A..29..195C} Combes, F.\ 1991, \araa, 29, 
195 

\bibitem[Freedman et al.(1994)]{1994ApJ...427..628F} Freedman, W.~L., et 
al.\ 1994, \apj, 427, 628 

\bibitem[Hollenbach \& Tielens(1999)]{1999RvMP...71..173H} Hollenbach, 
D.~J., \& Tielens, A.~G.~G.~M.\ 1999, Reviews of Modern Physics, 71, 173 

\bibitem[Kaufman et al.(1989)]{1989ApJ...345..697K} Kaufman, M., Elmegreen, 
D.~M., \& Bash, F.~N.\ 1989a, \apj, 345, 697 
 
\bibitem[Kaufman et al.(1989)]{1989ApJ...345..674K} Kaufman, M., Bash, 
F.~N., Hine, B., Rots, A.~H., Elmegreen, D.~M., \& Hodge, P.~W.\ 1989b, 
\apj, 345, 674 

\bibitem[R{\"o}llig et al.(2006)]{2006A&A...451..917R} R{\"o}llig, M., 
Ossenkopf, V., Jeyakumar, S., Stutzki, J., \& Sternberg, A.\ 2006, \aap, 
451, 917 

\bibitem[Sage \& Westpfahl(1991)]{1991A&A...242..371S} Sage, L.~J., \& 
Westpfahl, D.~J.\ 1991, \aap, 242, 371 
 
\bibitem[Smith et al.(2000)]{2000ApJ...538..608S} Smith, D.~A., Allen, 
R.~J., Bohlin, R.~C., Nicholson, N., \& Stecher, T.~P.\ 2000, \apj, 538, 
608 

\bibitem[Sternberg(1988)]{1988ApJ...332..400S} Sternberg, A.\ 1988, \apj, 
332, 400 

\bibitem[Sunada et al.(1995)]{1995ASPC...75..230S} Sunada, K., Noguchi, T., 
Tsuboi, M., \& Inatani, J.\ 1995, ASP Conf.~Ser., 75, 230 

\bibitem[Taylor and Wilson, 1998]{tay98} Taylor, C. L., \& Wilson, C.
D. 1998, \apj, 494, 581

\bibitem[Tielens(2005)]{2005pcim.book.....T} Tielens, A.~G.~G.~M.\
2005, The Physics and Chemistry of the Interstellar Medium (Cambridge:
Cambridge University Press)

\bibitem[Tilanus \& Allen(1987)]{1987sfig.conf..309T} Tilanus,
R.~P.~J., \& Allen, R.~J.\ 1987, in Star Formation in Galaxies,
ed. C. J. Lonsdale Persson (NASA Conf. Pub. 2466), 309

\bibitem[Ulich \& Haas(1976)]{1976ApJS...30..247U} Ulich, B.~L., \& Haas, 
R.~W.\ 1976, \apjs, 30, 247 

\bibitem[Visser(1980)]{1980A&A....88..159V} Visser, H.~C.~D.\ 1980, \aap, 
88, 159 

\bibitem[Williams \& Maddalena(1996)]{1996ApJ...464..247W} Williams, J.~P., 
\& Maddalena, R.~J.\ 1996, \apj, 464, 247 (WM96)

\bibitem[Young \& Scoville(1991)]{1991ARA&A..29..581Y} Young, J.~S., \& 
Scoville, N.~Z.\ 1991, \araa, 29, 581 

\end{thebibliography}
\end{document}